\documentclass[11pt]{article}

\usepackage{amsmath,amssymb}
\usepackage{latexsym}
\usepackage{bm}
\usepackage{graphicx}

\newtheorem{thm}{Theorem}

\newtheorem{lem}[thm]{Lemma}

\newenvironment{proof}{\begin{trivlist}
                       \item[]{\bf Proof.}
                       \hspace{0cm}}{\hfill $\Box$
                       \end{trivlist}}

\def\Image{\mathop{\rm Im}}
\def\supp{\mathop{\rm supp}}
\def\diam{\mathop{\rm diam}}

\begin{document}

\markboth{A. G. Ramm}{Electromagnetic wave scattering by small bodies}

\title{
Electromagnetic wave scattering by small bodies}

\author{A. G. Ramm$\dag$\\
\\
$\dag$Mathematics Department, Kansas State University,\\
Manhattan, KS 66506-2602, USA\\
email: ramm@math.ksu.edu\\
}

\date{}
\maketitle

\begin{abstract} \noindent A reduction of the Maxwell's system to a Fredholm
second-kind integral equation with weakly singular kernel is given
for electromagnetic (EM) wave scattering by one and many small bodies. 
This equation is solved asymptotically as the characteristic size of the 
bodies tends to zero. The
technique developed is used for solving the many-body 
EM wave scattering problem by rigorously reducing it to solving linear 
algebraic systems, completely bypassing the usage of integral 
equations. An
equation is derived for the effective field in the medium, in which many small
particles are embedded. A method for creating a desired refraction 
coefficient is outlined.

\footnote{key words:  electromagnetic waves, wave
scattering by small bodies,  many-body scattering, "smart" materials
}

\end{abstract}

\section{Introduction\label{s1}} There is a large literature on electromagnetic
wave (EM) scattering (\cite{[C]}- \cite{[M]}), to 
name a few books.
In \cite{[R476]}(see also \cite{[R117]}, \cite{[R190]} and references 
therein), wave
scattering theory is developed for small bodies of arbitrary shapes. The ideas
and methods from \cite{[R476]} were used in recent papers \cite{[R509]},
\cite{[R515]}, where a many-body scattering problem was solved asymptotically
for scalar wave scattering, as the characteristic size $a$ of small bodies
tends to zero. One of the applications of this theory in acoustics is a method
for creating new materials with the desired refraction coefficients. In
particular, the new materials may have a desired wave-focusing property: a
plane wave incident at a fixed direction with a fixed wavenumber will be
scattered by the new material with a desired radiation pattern. Another
application (\cite{[R533]}, (\cite{[R536]})) is a method for creating materials
with negative refraction,
that is, materials
in which the group velocity is directed opposite to the phase velocity.

In \cite{[R533]} the new materials are created by embedding small particles 
into a given
material in a bounded domain $D$ so that the number of the particles embedded
into a small subcube $\triangle\subset D$, is given by the formula
\begin{equation} \label{eq1}
\mathcal{N}(\triangle)=\frac{1}{a^{2-\varkappa}}\int_\triangle
N(x)dx[1+o(1)], \quad a\to 0, \end{equation} 
where $a$ is the radius
of the embedded spherical particles $D_m=B(x_m,a),$ $x_m$ is the center of the
ball $B(x_m,a)$, $a$ is its radius,  $0<\varkappa\leq 1$ is a 
parameter,
$N(x)\geq 0$ is an arbitrary continuous in $D$ function, which we can choose as 
we
want,  the boundary condition on the surface $S_m$ of the m-th particle
$D_m$ is: \begin{equation} \label{eq2}
\frac{\partial u}{\partial N}=\frac{h(x_m)}{a^\varkappa}u \quad \text{on} 
\
S_m, \end{equation} where $N$ is the unit normal to $S_m$, pointing out of
$D_m$, and $h(x)$ is an arbitrary continuous function which we can choose 
as we
want.

The purpose of this paper is to develop a 
theory for the electromagnetic (EM) wave scattering, similar to the one 
developed in 
\cite{[R509]}-\cite{[R516]}, \cite{[R536]}, \cite{[R539]}, in the sense 
that 
it yields a method for
creating materials with a desired refraction coefficient. Condition
\eqref{eq2} is not used in this paper, but formula \eqref{eq73}
of this paper is a generalization of \eqref{eq1}.

In Section \ref{s2} we present a theory for EM wave scattering by a single small 
body. In Section \ref{s3} many-body scattering
problem is solved. In Section \ref{s4} the negative refraction is 
discussed. In  Section \ref{s5} the conclusions are formulated. 

\section{EM wave scattering by a single small body\label{s2}}
Let $D$ be the body and $a:=\frac{1}{2}\diam  D.$ Assume that 
\begin{equation}
\label{eq3}
ka\ll 1,
\end{equation}
where $k>0$ is the wavenumber. The governing equations are:
\begin{equation}
\label{eq4}
\nabla\times E=i\omega\mu H, \quad \nabla\times 
H=-i\omega\epsilon'(x)E \quad \text{in} \, \mathbb{R}^3,
\end{equation}
where $\omega>0$ is the frequency, $\mu=\text{const}$ is the magnetic constant, 
$\epsilon'(x)=\epsilon>0$
in $D'=\mathbb{R}^3\setminus D, \quad 
\epsilon'(x)=\epsilon(x)+i\frac{\sigma(x)}{\omega};
\, \sigma(x)\geq 0, \, \epsilon'(x)\not=0 \, \forall x \in \mathbb{R}^3,\, \epsilon'(x) \in C^2(\mathbb{R}^3)$
is a twice continuously differentiable function, and $\sigma(x)=0$ in $D'$ is 
the conductivity. 
From (4) one gets
\begin{equation}
\label{eq5}
\nabla\times\nabla\times E=K^2(x)E, \quad H=\frac{\nabla\times E}{i\omega\mu},
\end{equation}
\begin{equation}
\label{eq6}
K^2(x):=\omega^2 \epsilon'(x) \mu.
\end{equation}

We seek the solution of the equation
\begin{equation}
\label{eq7}
\nabla\times\nabla\times E=K^2(x)E
\end{equation}
satisfying the radiation condition: 
\begin{equation}
\label{eq8}
E(x)=E_0(x) + v,
\end{equation}
where $E_0(x)$ is the plane wave,
\begin{equation}
\label{eq9}
E_0(x)=\mathcal{E} e^{ik\alpha\cdot x}, \quad  
k=\frac{\omega}{c},
\end{equation}
where $c=\omega\sqrt{\epsilon\mu}$ is the wave velocity in the homogeneous medium 
outside $D$, $\epsilon=const$ is the dielectric parameter in the outside 
region $D'$, $\mu=const$,
$\alpha\in S^2$ is the incident direction of the plane wave, 
$\mathcal{E}\cdot\alpha=0,$ $\mathcal{E}$ is a constant vector,
and the scattered field $v$ satisfies the radiation condition
\begin{equation}
\label{eq10}
\frac{\partial v}{\partial r}-ikv=o\big(\frac{1}{r}\big),\quad r=|x|\to\infty,
\end{equation}
uniformly in directions $\beta:=\frac{x}{r}.$

If $E$ is found, then the pair $\{E,H\}$, where $H=\frac{\nabla\times E}{i\omega\mu},$ 
solves our scattering problem. 

Our goal is to derive a Fredholm second-kind integral equation for $E$. 

Let us rewrite equation \eqref{eq7} as 
\begin{equation}
\label{eq11}
-\triangle E+\nabla(\nabla\cdot E)-k^2E-p(x)E=0,
\end{equation}
where
\begin{equation}
\label{eq12}
p(x):=K^2(x)-k^2, \quad p(x)=0 \, \text{in}\ D'.
\end{equation}
Note that $\Image K^2(x)=\Image p(x)\geq 0$. This will be used
in the proof of Claim 3 below (see \eqref{eq23}).

It follows from \eqref{eq7} that 
\begin{equation}
\label{eq13}
0=\nabla\cdot(K^2(x)E)=\nabla K^2(x)\cdot E+K^2(x)\nabla\cdot E.
\end{equation}
From \eqref{eq12} and \eqref{eq13} one gets
\begin{equation}
\label{eq14}
-\triangle E-k^2E-p(x)E-\nabla(q(x)\cdot E)=0,
\end{equation}
where
\begin{equation}
\label{eq15}
q(x):=\frac{\nabla K^2(x)}{K^2(x)}, \quad q(x)=0 \, \text{in}\ D'.
\end{equation}

From \eqref{eq14} and \eqref{eq8} one gets
\begin{equation}
\label{eq16}
E=E_0+\int_Dg(x,y)\bigg{(}p(y)E(y)+\nabla_y(q(y)\cdot E(y))\bigg{)}dy,
\end{equation}
where
\begin{equation}
\label{eq17}
g(x,y):=\frac{e^{ik|x-y|}}{4\pi|x-y|}.
\end{equation}
If $q$ vanishes at the boundary of $D$, i.e., $\nabla K^2(x)=\nabla 
p(x)$ vanishes at the boundary of $D$, then, after 
an integration 
by parts in the last term of \eqref{eq16}, and taking into account that 
$\nabla_x g(x,y)=-\nabla_y g(x,y)$,
one gets an equivalent 
equation, which we also refer to as \eqref{eq16}:
$$E=E_0+\int_Dg(x,y)p(y)E(y)dy +\nabla_x\int_Dg(x,y)q(y)\cdot 
E(y)dy.
$$

Since we have assumed $K^2(x)\not=0$ and $K^2(x)\in C^2(\mathbb{R}^3)$, it 
follows that 
equation \eqref{eq16} is a Fredholm equation of the second kind, because 
the integral operator $T$
in \eqref{eq16} is compact in $H^1(D)$, where $H^1(D)$ is the usual Sobolev 
space.

Indeed, the operator $Bf:=\int_Dg(x,y)f(y)dy$ acts from $L^2(D)$ into $H^2(D)$,
and the operator $B_1E:=p(y)E+\nabla\big{(}q(y)\cdot E\big{)}$ acts from
$H^1(D)$ into $L^2(D)$.  Thus, $T$ acts from $H^1(D)$ into $H^2(D)$ and is,
therefore, compact by the embedding theorem.

Let us summarize the result.
\begin{lem}
\label{Lemma1}
If $K^2(x)\in C^2(\mathbb{R}^3),\ K^2(x)\not=0,\ K^2(x)=k^2>0$ in the exterior 
domain 
$D':=\mathbb{R}^3\setminus D$,  then the operator $T$
in \eqref{eq16} is compact in $H^1(D)$, so that equation \eqref{eq16} is of 
Fredholm type in $H^1(D).$
\end{lem}

\begin{lem}
\label{Lemma2}
Equation \eqref{eq16} is uniquely solvable in $H^1(D)$.
\end{lem}

\begin{proof} Let us prove Lemma 2. It is sufficient to prove that the 
homogeneous version of
equation \eqref{eq16} has only the trivial solution. If $E$ solves the
homogeneous equation \eqref{eq16}, then $E$ solves equation \eqref{eq7} and
satisfies the radiation condition \eqref{eq10}. The only solution to
\eqref{eq7} satisfying \eqref{eq10} is the trivial solution $E=0$.

Let us give details and prove the above claims.

\underline{Claim 1}.\,
A solution to \eqref{eq16} satisfies equation \eqref{eq7}, and, 
consequently, \eqref{eq13}, that is, $\nabla\cdot(K^2(x)E)=0$.

Thus, Claim 1 states that \eqref{eq16} is equivalent to the original
equation \eqref{eq7}, which is not at all obvious.

\underline{Claim 2}.\, 
The only solution to \eqref{eq7}, \eqref{eq10} is $E=0$.

\underline{Proof of Claim 1}.\,
If $E$ solves \eqref{eq16} and is of the form \eqref{eq8}, then 
$$
(-\triangle-k^2)E=p(x)E+\nabla(q(x)\cdot E).
$$

Rewrite this equation as
$$
\nabla\times\nabla\times E-\nabla(\nabla\cdot E)-\nabla(q(x)\cdot 
E)=K^2(x)E,
$$
or
\begin{equation}
\label{eq18}
\nabla\times\nabla\times E-\nabla\bigg{(}\frac{K^2(x)\nabla\cdot E+\nabla K^2(x)\cdot E}{K^2(x)}\bigg{)}
=K^2(x)E
\end{equation}
Denote 
$$\frac{1}{K^2(x)}\nabla\cdot(K^2(x)E):=\psi(x).$$ 
Taking divergence of equation \eqref{eq18}
one gets
\begin{equation}
\label{eq19}
-\triangle\psi-K^2(x)\psi=0 \quad \text{in}\,\quad \mathbb{R}^3.
\end{equation}
The function $\psi$ satisfies the radiation condition \eqref{eq10}.

\underline{Claim 3}.\,
The only solution to \eqref{eq19}, which satisfies condition \eqref{eq10}, 
is $\psi=0$.

We prove Claim 3 below. Assuming that this claim is proved, we infer that $\psi=0$,
so $\nabla\cdot(K^2(x)E)=0$, and equation \eqref{eq7} holds.

Claim 1 is proved.      $\Box$ 

\underline{Let us prove Claim 3}.\, Equation \eqref{eq19} can be written as:
\begin{equation}
\label{eq20}
-\triangle\psi-k^2\psi-p(x)\psi=0 \quad \text{in} \, \mathbb{R}^3,
\end{equation}
where $p(x)$ is defined in equation \eqref{eq12}, and $p(x)=0$ in $D'$. 
It is known that if $k^2>0,\, p(x)\in L^2(D)$,
$\Image p(x)\geq 0$, $D$ is a bounded domain, and 
$\psi$ satisfies the 
radiation condition \eqref{eq10} and equation \eqref{eq20}, then $\psi=0$ 
(see, e.g., \cite{[R509]}). 

For convenience of the reader, we sketch the proof. From \eqref{eq20} and its
complex conjugate one derives the relation 
\begin{equation} \label{eq21}
\psi\triangle\bar{\psi}-\bar{\psi}\triangle\psi-2i\Image p(x)|\psi|^2=0.
\end{equation} 
Integrate \eqref{eq21} over a ball $B_R$ centered at the origin
of radius $R$, and use the Green's formula to get 
\begin{equation}
\label{eq22} \int_{|x|=R}\big{(}\psi\frac{\partial\bar{\psi}}{\partial
r}-\bar{\psi}\frac{\partial\psi}{\partial r}\big{)}ds -2i\int_D \Image
p(x)|\psi|^2dx=0. 
\end{equation} 
By the radiation condition \eqref{eq10} for
$\psi$ one can rewrite \eqref{eq22} as 
\begin{equation} \label{eq23}
-2ik\lim_{R\to\infty}\int_{|x|=R}|\psi|^2ds-2i\int_D\Image p(x)|\psi|^2dx=0.
\end{equation} 
Since $\Image p(x)\geq 0$ by our assumptions, and $k>0$, it
follows that 
\begin{equation} \label{eq24}
\lim_{R\to\infty}\int_{|x|=R}|\psi|^2dx=0. 
\end{equation} 
Relation \eqref{eq24}
and the equation 
\begin{equation} 
\label{eq25} (\triangle+k^2)\psi=0 \quad |x|>
R_0, 
\end{equation} 
where $B_{R_0}\supset D$, imply $\psi=0$ for $|x|>R_0.$
(See \cite[p.25]{[R190]}).

By the unique continuation principle for the solutions of the homogeneous 
Schr\"odinger equation
\eqref{eq20}, it folows that $\psi=0$. Claim 3 is proved. $\Box$ 

\underline{Let us prove Claim 2}.\, This will complete the proof of 
Lemma \ref{Lemma2}.

If $E$ solves \eqref{eq7} and satisfies \eqref{eq10}, the the pair $\{E,H\}$,
where $H$ is defined by the second formula in \eqref{eq5}, solves the
homogeneous Maxwell system \eqref{eq4}, satisfies the radiation condition, and
$\{E,H\}$ is in $H_{loc}^2(\mathbb{R}^3)$. It is known (see, e.g., \cite{[Mu]})
that this implies $E=H=0.$ Lemma \ref{Lemma2} is proved.  \end{proof}

From formula \eqref{eq16} assuming that $a:=\frac{1}{2}\diam D$ is small,
$ka\ll 1$ and the origin is inside $D$, one gets: 
\begin{equation} \label{eq26}
E=E_0+\frac{e^{ik|x|}}{|x|}\frac{1}{4\pi}\int_De^{-ik\beta\cdot y}
\big{(}p(y)E(y)+\nabla(q(y)\cdot E)\big{)}dy[1+O(\frac{a}{|x|})] 
\end{equation}
as $|x|\to\infty,\, \frac{x}{|x|}:=\beta$, and \begin{equation} \label{eq27}
e^{-ik\beta\cdot y}=1+O(ka), \quad |y|\leq a. \end{equation} 
Thus, 
\begin{equation} \label{eq28}
E(x)=E_0(x)+\frac{e^{ik|x|}}{|x|}\frac{1}{4\pi}\int_D
\big{(}p(y)E(y)+\nabla(q(y)\cdot E)\big{)}dy[1+O(\frac{a}{|x|})].
\end{equation} 
Here we have used the estimate $$ (|x|^2-2x\cdot
y+|y|^2)^{\frac{1}{2}} =|x|-\beta \cdot y+O(\frac{|y|^2}{|x|}), \quad 
|x|\gg a\geq |y|, \quad \beta:=x/|x|, $$ 
which yields $$ \frac{e^{ik|x-y|}}{4\pi|x-y|}=
\frac{e^{ik|x|-ik\beta\cdot
y+O(\frac{ka^2}{|x|})}}{4\pi|x|\big{(}1+O(\frac{a}{|x|})\big{)}}
=\frac{e^{ik|x|}}{4\pi|x|}e^{-ik\beta\cdot
y}\big[1+O(\frac{a}{|x|})+O(\frac{ka^2}{|x|})\big] $$ where $|y|\leq 
a\ll|x|$. If
$ka\ll 1$, then $O(\frac{ka^2}{|x|})\ll O(\frac{a}{|x|})$, and we get estimate
\eqref{eq28}. Usually the error term in \eqref{eq28} is written as
$O(\frac{1}{|x|})$ as $|x|\to\infty,$ but for small $a$ the term
$O(\frac{a}{|x|})$ is small already for $|x|\gg a$, and $x$ may be still small.

If $D$ does not contain the origin and $x_m\in D$, then one can rewrite 
\eqref{eq16} as:
\begin{equation}
\label{eq29}
\begin{split}
E(x)=&E_0(x)+g(x,x_m)\int_Dp(y)E(y)dy+\nabla_xg(x,x_m)\int_D q(y)\cdot E(y)dy\\
&+\int_D\big{[}g(x,y)-g(x,x_m)\big{]}p(y)E(y)dy +\nabla_x\int_D\big{[}g(x,y)-
g(x,x_m)\big{]}q(y)\cdot E(y)dy,
\end{split}
\end{equation}
where $g$ is defined in \eqref{eq17}.
Let 
\begin{equation}
\label{eq30}
V_m:=\int_Dp(y)E(y)dy, \quad \nu_m=\int_Dq(y)\cdot E(y)dy,
\end{equation}
where $p(x)$ is defined in \eqref{eq12} and $q(x)$ is defined in 
\eqref{eq15}.

If 
$$d:=|x-x_m|\gg a, \, a:=\frac{1}{2}\diam  D,$$ 
then
\begin{equation}
\label{eq31}
\big{|}g(x,y)-g(x,x_m)\big{|}=\big{|}\nabla_yg(x,\tilde{y})\cdot(\tilde{y}-x_m)\big{|}
\leq ca\max(\frac{k}{d},\frac{1}{d^2}),
\end{equation}
where $c>0$ stand for various constants independent of $a$, 
$\tilde{y}$ is an intermediate point, and $d=|x-x_m|$. Thus,
$$
J_m:=\bigg{|}\int_D[g(x,y)-g(x,x_m)]p(y)E(y)dy\bigg{|}
\leq ca\max(\frac{k}{d},\frac{1}{d^2})|p(x_m)E(x_m)||D|,
$$
where $|D|=O(a^3)$ is the volume of $D$, and we assume that $p(x)$ and $E(x)$ 
are continuous in $D$.

On the other hand, 
\begin{equation}
\label{eq32}
I_m:=\bigg{|}g(x,x_m)\int_Dp(y)E(y)dy\bigg{|}=O\big{(}\frac{|p(x_m)E(x_m)||D|}{d}
\big{)}\gg J_m,
\end{equation}
provided that $\frac{a}{d}\ll 1$ and $ka\ll 1$. These assumptions hold 
throughout the paper. 

We have checked that
$$J_m\ll I_m.$$

Similarly,
\begin{equation}
\label{eq33}
J'_m \ll I'_m, \quad a\ll d, \quad ka\ll 1,
\end{equation}
where
\begin{equation}
\label{eq34}
I'_m:=\bigg{|}\nabla_xg(x,x_m)\int_Dq(y)\cdot E(y)dy\bigg{|},
\end{equation}
\begin{equation}
\label{eq35}
J'_m:=\bigg{|}\nabla_x\int_D[g(x,y)-g(x,x_m)]q(y)\cdot E(y)dy\bigg{|}.
\end{equation}
Also, 
$$J_m\ll I'_m, \quad J'_m\ll I_m.$$ 

Therefore, in the region $|x-x_m|\gg a$ one can calculate $E(x)$ by the approximate formula
\begin{equation}
\label{eq36}
E(x)=E_0(x)+g(x,x_m)V_m+\nabla_xg(x,x_m)\nu_m,
\end{equation}
with the error $O(\frac{a}{d}+ka)$. Here $V_m$ and $\nu_m$ are defined in 
\eqref{eq30}.

{\it Let us derive formulas for the vector $V_m$ and the scalar $\nu_m$.}

Multiply \eqref{eq29} by $p(x)$ and integrate over $D$ to get:
\begin{equation}
\label{eq37}
V_m=V_{0m}+\int_Dp(x)g(x,x_m)dx\,V_m+\int_Dp(x)\nabla_xg(x,x_m)dx\,\nu_m,
\end{equation}
where 
$$V_{0m}:=\int_Dp(x)E_0(x)dx,$$ 
and we have neglected the two last terms in \eqref{eq29}. 
These terms are smaller that the ones we kept, as we prove soon. 

Let
\begin{equation}
\label{eq38}
\int_Dp(x)g(x,x_m)dx:=a_m, \quad \int_Dp(x)\nabla_xg(x,x_m)dx:=A_m.
\end{equation}
Take dot product of \eqref{eq29} with $q(x)$, integrate over $D$ and 
again neglect two last terms to get
\begin{equation}
\label{eq39}
\nu_m=\nu_{0m}+B_m\cdot V_m+b_m\nu_m,
\end{equation}
where
\begin{equation}
\label{eq40}
\nu_{0m}:=\int_Dq(x)\cdot E_0(x)dx,
\end{equation}
\begin{equation}
\label{eq41}
B_m:=\int_Dq(x)g(x,x_m)dx, \quad b_m:=\int_Dq(x)\cdot\nabla_xg(x,x_m)dx.
\end{equation}
From \eqref{eq37} one gets
\begin{equation}
\label{eq42}
V_m=\frac{V_{0m}}{1-a_m}+\frac{A_m}{1-a_m}\nu_m.
\end{equation}
Note that if the body $D$ is small, then $|a_m|<1$,  
and actually $|a_m|\ll 1$, so \eqref{eq42}
is well defined. Also,  $|b_m|<1$ if $D$ is small.
From \eqref{eq42} and \eqref{eq39} one obtains:
\begin{equation}
\label{eq43}
\nu_m=\frac{(1-a_m)\nu_{0m}+B_m\cdot V_{0m}}{(1-a_m)(1-b_m)-B_m\cdot A_m}.
\end{equation}
and
\begin{equation}
\label{eq44}
V_m=\frac{V_{0m}}{1-a_m}+\frac{A_m}{1-a_m}\cdot
\frac{(1-a_m)\nu_{0m}+B_m\cdot V_{0m}}{(1-a_m)(1-b_m)-B_m\cdot A_m}.
\end{equation}
{\it Formulas \eqref{eq36}, \eqref{eq43}, and \eqref{eq44} solve the 
scattering problem of EM waves
by one small body $D$ located so that $x_m\in D$.}

Let us now check that one may neglect the terms we have neglected. 

For example, one has
\begin{equation*}
\begin{split}
\int_Ddxp(x)\int_D\big{[}g(x,y)-g(x,x_m)\big{]}p(y)E(y)dy
&=\int_Ddyp(y)E(y)\int_Ddxp(x)\big{[}g(x,y)-g(x,x_m)\big{]}\\
&=\int_Ddyp(y)E(y)\big{[}\psi_1(y)-\psi_1(x_m)\big{]}\\
&\ll \int_Ddy\big{|}p(y)E(y)\big{|},
\end{split}
\end{equation*}
because that $\psi_1(y):=\int_Ddxp(x)g(x,y)$ is a continuous function. 

The function $\psi_1(y)$ is  continuous if $p(x)\in L^2(D)$, since 
$\psi_1\in H_{loc}^2(\mathbb{R}^3)$
and the embedding from $H^2(D)$ into $C(D)$ is continuous (and compact if 
$D\subset\mathbb{R}^3$ is a bounded domain, which we have assumed).

A similar argument holds for 
$\psi_2(y):=\int_D\big{[}\nabla_xg(x,y)-\nabla_x(g(x,x_m)\big{]}\cdot q(y)dy$
and justifies the possibility to neglect the second term we have neglected.

To summarize: 

{\it If $p(x)$ and $q(x)$ are continuous functions then the 
scattering of EM waves
by a small body $D$ is solved analytically by the formulas \eqref{eq36}, 
\eqref{eq43}, \eqref{eq44},
with an error of order $O\big{(}\frac{a}{d}+ka\big{)}$.}

\section{EM wave scattering by many small bodies. Creating materials with a 
desired radiation coefficient.\label{s3}}
The basic equation now has the form
\begin{equation}
\label{eq45}
\begin{split}
E(x)&=E_0(x)+\sum_{m=1}^M\int_{D_m}g(x,y)p(y)E(y)dy+\sum_{m=1}^M\nabla_x
\int_{D_m}g(x,y)q(y)\cdot E(y)dy\\
&=E_0(x)+\sum_{m=1}^M\big{[}g(x,x_m)V_m+\nabla_xg(x,x_m)\nu_m\big{]}+
\sum_{m=1}^M(J_m+K_m),
\end{split}
\end{equation}
where $M$ is the number of small bodies,
\begin{equation}
\label{eq46}
J_m:=\int_{D_m}\big{[}g(x,y)-g(x,x_m)\big{]}p(y)E(y)dy,
\end{equation}
and
\begin{equation}
\label{eq47}
K_m:=\nabla_x\int_{D_m}\big{[}g(x,y)-g(x,x_m)\big{]}q(y)\cdot E(y)dy.
\end{equation}
We assume throughout, that the points $x_m\in D_m$ are distributed in $D$ 
so that their number $\mathcal{N}(\triangle)$ in any subdomain $\triangle 
\in D$ of $D$ is given by formula \eqref{eq73} (see below) with
$\varphi(a)=a^{3-\varkappa},\,\varkappa>0$ (see Lemma 3 below).
As $a\to 0$ and $M\to \infty$, the set of the points $x_m$ becomes 
everywhere dense in $D$.

Neglecting $J_m$ and $K_m$, let us derive a linear algebraic system (las) for 
finding $V_m$ and $\nu_m$.
If $V_m$ and $\nu_m, \, 1\leq m\leq M$, are found, then the EM 
wave scattering problem for $M$
small bodies is solved by the formula
\begin{equation}
\label{eq48}
E(x)=E_0(x)+\sum_{m=1}^M\big{[}g(x,x_m)V_m+\nabla_xg(x,x_m)\nu_m\big{]}
\end{equation}
with an error $O\big{(}\frac{a}{d}+ka\big{)}$ in the region 
$\min_{1\leq m\leq M}|x-x_m|:=d\gg a$.

To derive a linear algebraic system for $V_m$ and $\nu_m$, multiply 
\eqref{eq45} by $p(x)$, integrate over $D_j$, 
and neglect the terms $J_m$ and $K_m$, to get:
\begin{equation}
\label{eq49}
V_j=V_{0j}+\sum_{m=1}^M\big{(}a_{jm}V_m+B_{jm}\nu_m\big{)}, \quad 1\leq j\leq M,
\end{equation}
where
\begin{equation}
\label{eq50}
V_{0j}:=\int_{D_j}p(x)E_0(x)dx,\quad a_{jm}=\int_{D_j}p(x)g(x,x_m)dx,\quad 
B_{jm}:=\int_{D_j}p(x)\nabla_xg(x,x_m)dx.
\end{equation}
Take the dot product of \eqref{eq45} with $q(x)$, integrate over $D_j$, and neglect $K_m,\,J_m$, to get:
\begin{equation}
\label{eq51}
\nu_j=\nu_{0j}+\sum_{m=1}^M\big{(}C_{jm}V_m+d_{jm}\nu_m\big{)}, 
\end{equation}
where
\begin{equation}
\label{eq52}
\nu_{0j}=\int_{D_j}q(x)\cdot E_0(x)dx,\quad C_{jm}:=\int_{Dj}q(x)g(x,x_m)dx,
\quad d_{jm}:=\int_{D_j}q(x)\cdot\nabla_xg(x,x_m)dx.
\end{equation}
Equations \eqref{eq49} and \eqref{eq51} form a linear algebraic system for 
finding $V_m$ and $\nu_m,\,1\leq m\leq M.$
This linear algebraic system is uniquely solvable if $ka\ll 1$ and $a\ll d$. 
The elements $C_{jm}$ and $B_{jm}$
are vectors, and $a_{jm}$, $d_{jm}$ are scalars. Under the conditions
\begin{equation}
\label{eq53}
\max_{1\leq j\leq 
M}\sum_{m=1}^M\big{(}|a_{jm}|+|d_{jm}|+\|B_{jm}\|+\|C_{jm}\|\big{)}<1,
\end{equation}
one can solve linear algebraic system \eqref{eq49} and \eqref{eq51} by 
iterations. In \eqref{eq53} $\|B_{jm}\|$ and $\|C_{jm}\|$
are the lengths of the corresponding vectors. Condition \eqref{eq53} 
holds if $a\ll 1$ and $M$ is not growing too fast as $a\to 0$,
not faster than $O(a^{-3})$.

Consider the limit, as $a\to 0$ and $M\to\infty$,
of the field $E(x):=E(x,a)$. Let us assume that the number 
$\mathcal{N}(\triangle)$
of small particles in any open subset $\triangle\subset D$ is given by the formula:
\begin{equation}
\label{eq54}
\mathcal{N}(\triangle)=\frac{1}{a^{3\varkappa_1}}\int_\triangle N(x)dx\big{[}1+o(1)\big{]},\,a\to 0
\end{equation}
where $0<\varkappa_1<1$ is some number, and 
$N(x)\geq 0$ is a
continuous in $D$ function. If \eqref{eq54} holds, then the distance 
between two neighboring small particles is of the order 
$O(a^{\varkappa_1})$. We choose $\varkappa_1=(3-\varkappa)/3$ above 
formula \eqref{eq75}.

Let us write \eqref{eq45} as
\begin{equation}
\label{eq55}
E(x)=E_0(x)+\sum_{m=1}^M\bigg{[}g(x,x_m)\int_{D_m}p(y)dyE(x_m)+
\nabla_xg(x,x_m)\int_{D_m}\frac{\nabla_yK^2(y)}{K^2(y)}\cdot E(y)dy\bigg{]},
\end{equation}
where the terms $J_m$ and $K_m$ are neglected. 

Choose the function $p$ so that $\supp p=\bigcup_{m=1}^MB_m$, where
$D_m:=B_m:=\{y:|y-x_m|\leq a\}$, that is, the small particles are balls
of radius $a$ centered at the points $x_m$, and $\supp p$ is the support of 
$p$. 
Choose $p$ of the form:
\begin{equation}
\label{eq56}
p(y)=p(r,a),\quad y\in B_m,\quad r:=|y-x_m|,\, 1\leq m\leq M,
\end{equation}
where the function $p(r,a)$ should be chosen so that
\begin{equation}
\label{eq57}
\int_{B_m}p(y)dy=O(a^s),\quad a\to 0,\quad s:=3\varkappa_1=3-\varkappa,
\end{equation}
where $0<\varkappa<3$.
One may like to have $Z_m=O(a^s)$ as $a\to 0$, where
\begin{equation}
\label{eq58}
Z_m:=\int_{B_m}\frac{\nabla p(y)\cdot E(y)}{k^2+p(y)}dy,
\end{equation}
with the same $s$ as in \eqref{eq58}. If this would be possible, then 
it would make possible passing to the limit as $a\to 0$ similarly to the
passing done in \cite{[R536]}.
However, as we will see below, the relations \eqref{eq57} and $Z_m=O(a^s)$ 
are not 
compatible, in general. 

One has:
\begin{equation}
\label{eq59}
\nabla p(y)=p'(r,a) r^0,\quad r^0:=\frac{y-x_m}{r}, \quad \quad 
p':=\frac{dp}{dr}.
\end{equation}
Here $r^0$ is a unit vector and $r$ is the length of the vector $y-x_m$.
Let
\begin{equation}
\label{eq60}
E(y)=E(x_m)+re_j\frac{\partial E_j(x_m)}{\partial y_i}r^0_i+O(a^2),
\quad r^0_i:=\frac{y_i-(x_m)_i}{r},
\end{equation}
$(x_m)_i:=x_m\cdot e_i$, $a\cdot b$ is the dot product of vectors $a$ and 
$b$,
$e_i\cdot e_j=\delta_{ij}$ is the standard Cartesian basis of $\mathbb{R}^3$, and 
$\delta_{ij}=
\left\{
\begin{matrix}
&0,\,i\not= j,\\
&1,\, i=j
\end{matrix}
\right.$. 

Over the repeated indices in \eqref{eq60} and everywhere below summation is understood. 

If $p(y)=p(r,a)$ then
\begin{equation}
\label{eq61}
\int_{B_m}\frac{\nabla p(y)\cdot E(x_m)}{k^2+p(y)}dy=0
\end{equation}
for any constant vector $E(x_m)$ by symmetry.

Thus 
\begin{equation}
\label{eq62}
\int_{B_m}\frac{\nabla p(y)\cdot 
E}{k^2+p(y)}dy=\int_{B_m}\frac{p'(r,a)rr^0_jr^0_i}{k^2+p(r,a)}dy 
E_{j,i}(x_m),
\quad E_{j,i}(x_m):=\frac{\partial E_j (x_m)}{\partial 
y_i}.
\end{equation}
By symmetry the integral in \eqref{eq62} vanishes if $i\not=j$. This 
integral for $i=j$
is equal to
\begin{equation}
\label{eq63}
Y_i:=\int_0^a\frac{drr^2rp'(r,a)}{k^2+p(r,a)}\int_0^\pi d\theta\sin\theta
\int_0^{2\pi}d\varphi r_i^2,
\quad i=1,2,3, \quad r_i:=r_i^0.
\end{equation}
One has $r_3^2=\cos^2\theta,\quad r_2^2=\sin^2\theta\sin^2\varphi,
\quad r_1^2=\sin^2\theta\cos^2\varphi$,
so
$$\int_0^\pi d\theta\sin\theta\int_0^{2\pi}d\varphi 
r_i^2=\frac{4\pi}{3},$$
and, with $r=ta$, one gets:
\begin{equation}
\label{eq64}
Y_i=\frac{4\pi}{3}a^4\int_0^1\frac{dtt^3p'_r(at,a)}{k^2+p(at,a)}+o(a^4),\quad 
a\to 0,\quad i=1,2,3.
\end{equation}
Thus
\begin{equation}
\label{eq65}
Z_m\sim \frac{4\pi}{3}a^4 I(a)\nabla\cdot E(x_m), \quad 
I(a):=\int_0^1\frac{p'_r(at,a)t^3dt}{k^2+p(at,a)},\qquad r=ta,
\end{equation}
and
\begin{equation}
\label{eq66}
j_m:=\int_{B_m}pdy=4\pi a^3\int_0^1dtt^2p(at,a):=4\pi a^3j(a).
\end{equation}

Let
\begin{equation}
\label{eq67}
p(r)=p(r,a)=\left\{
\begin{matrix}
&\frac{\gamma_m}{4\pi a^\varkappa}(1-t)^2h(t),\,0\leq t\leq 1,\,
t:=\frac{r}{a},\,\varkappa=\text{const}>0,\\
&0,\quad t>1,
\end{matrix}
\right.
\end{equation}
where $\gamma_m,\,\Image\gamma_m\geq0$, is a number we can choose at will. Then
\begin{equation}
\label{eq68}
j_m:=\int_{B_m}p(y)dy=\frac{a^3}{a^\varkappa}\gamma_m\int_0^1(1-t)^2h(t)t^2dt:=c_{1m}a^{3-\varkappa},
\end{equation}
where $c_{1m}$ is proportional to $\gamma_m$, so we can also choose 
$c_{1m}$ at will. 
By \eqref{eq65},
\begin{equation}
\label{eq69}
I(a)=\int_0^1\frac{t^3\{\gamma_m(1-t)^2h(t)\}'_t}{\nu+\gamma_m(1-t)^2h(t)}dt,
\quad
\nu:=4\pi k^2a^\varkappa.
\end{equation}
Can one choose $h(t)$ so that $I(a)=Ca^{-\varkappa}[1+o(1)]$ as $a\to 0$?
If yes, then $Z_m$ and $j_m$ are of the same order as $a\to 0$.

In fact, {\it it is not possible to choose such an $h(t)$}. One has
\begin{equation}
\label{eq70}
I(a)=t^3\ln[\nu+\gamma_m(1-t)^2h(t)]\big{|}_0^1-3\int_0^1t^2\ln[\nu+\gamma_m(1-t)^2h(t)]dt.
\end{equation}
Thus
\begin{equation}
\label{eq71}
I(a)=\ln\nu+O(1)=\varkappa\ln a+O(1),\quad a\to 0.
\end{equation}

Therefore no choice of $h\in C[0,1]$ can ensure the asymptotics 
$I(a)=Ca^{-\varkappa}[1+o(1)]$
as $a\to 0$.

We have obtained the asymptotic formulas:
\begin{equation}
\label{eq72}
j_m=c_{1m}a^{3-\varkappa},
\quad Z_m=\frac{4\pi\varkappa}{3}(\nabla\cdot E)(x_m)a^3\ln a[1+o(1)],\quad 
a\to 0.
\end{equation}
Let us take $h(t)=1$. Then $c_{1m}=\gamma_m /30$.
To pass to the limit $a\to 0$ in formula \eqref{eq55} we use the following 
lemma, which a generalization of a lemma from 
\cite{[R509]}:
\begin{lem}
\label{Lemma31}
If $f\in C(D)$ and $x_m$ are distributed so that
\begin{equation}
\label{eq73}
\mathcal{N}(\triangle)=\frac{1}{\varphi(a)}\int_\triangle N(x)dx\big{[}1+o(1)\big{]},
\quad a\to 0,\, 
\end{equation}
for any subdomain $\triangle \subset D$, where $\varphi(a)\geq 0$ is a 
continuous, monotone, strictly growing function,  
$\varphi(0)=0$,
then
\begin{equation}
\label{eq74}
\lim_{a\to 0}\sum_mf(x_m)\varphi(a)=\int_Df(x)N(x)dx.
\end{equation}
\end{lem}

\begin{proof}
Let $D=\cup_{p}\triangle_p$ be a partition of $D$ into a union of small 
cubes 
$\triangle_p$, having no common interior points. Let
$|\triangle_p|$ denote the volume of $\triangle_p$, $\delta:=\max_p \diam 
\triangle_p$, 
and $y$ be the center of the cube $\triangle_p$.
One has
\begin{equation*}
\begin{split}
\lim_{a\to 0}\sum_mf(x_m)\varphi(a)
&=\lim_{\delta \to 0}\sum_{y^{(p)}\in\triangle_p}f(y^{(p)})\lim_{a\to 
0}\varphi(a)
\sum_{x_m\in\triangle_p}1\\
&=\lim_{\delta\to 0}\sum 
f(y^{(p)})N(y^{(p)})|\triangle_p|=\int_Df(x)N(x)dx.
\end{split}
\end{equation*}
The last equality holds since the preceding sum is a Riemannian sum
for the continuous function $f(x)N(x)$ in the bounded domain $D$.
Thus, Lemma 3 is proved.
\end{proof}

Let $\varphi(a)=a^{3-\varkappa},\,\varkappa>0$. Then, by formulas 
\eqref{eq58}, \eqref{eq65} 
and \eqref{eq72}, one gets:
\begin{equation}
\label{eq75}
\lim_{a\to 0}\frac{Z_m}{\varphi(a)}=0,\quad \lim_{a\to 
0}\frac{j_m}{\varphi(a)}=c_{1m}\not=0.
\end{equation}

Therefore, by Lemma \ref{Lemma31}, the right side of formula \eqref{eq55} has a 
limit as $a\to 0$
provided that \eqref{eq54} holds with $3\varkappa_1=3-\varkappa$, where 
$\varkappa$ is defined
in formula \eqref{eq68}. The limiting equation \eqref{eq55} takes the form
\begin{equation}
\label{eq76}
E_e(x)=E_0(x)+\int_Dg(x,y)C(y)E_e(y)dy,
\end{equation}
where $E_e(x):=\lim_{a\to 0}E(x)$ is the effective field in the medium
when the number of small particles tends to infinity while their size
$a$ tends to zero, 
and
\begin{equation}
\label{eq77}
C(x_m):=c_{1m} N(x_m).
\end{equation}

Formula \eqref{eq77} defines uniquely a continuous function $C(x)$ since the 
points $x_m$
are distributed everywhere dense in $D$ as $a\to 0$. The functions C(x) can be 
created at our
will, since it is defined by the numbers $c_{1m}$ and $N(x_m)$, which are at 
our disposal.


Apply the operator $\nabla^2+k^2$ to \eqref{eq76} and get 
\begin{equation}
\label{eq78}
[\nabla^2+\mathcal{K}^2(x)]E_e=0,\quad 
\mathcal{K}^2(x):=k^2+C(x):=k^2n^2(x).
\end{equation}
Therefore the (refraction) coefficient $n^2(x)$ is defined by the formula:
\begin{equation}
\label{eq79}
n^2(x)=1+k^{-2}C(x).
\end{equation}
The function $\mathcal{K}^2(x)$ is quite different from
the function $K^2(x)$, defined in \eqref{eq6}.
The function $C(x)$ and, therefore, $n^2(x)$ depend on 
the choice of $N(x)$ in \eqref{eq73} and on 
the choice of the coefficients $c_{1m}$
in  \eqref{eq72}. The $c_{1m}$ is determined by $\gamma_m$,
defined in \eqref{eq67}. {\it Since one is free to choose 
$\gamma_m$ and $N(x)$ as one wishes, one can create 
a desired function $n^2(x)$ by choosing suitable $\gamma_m$
and $N(x)$.}

Note that the total volume of the small particles is negligible as $a\to0$.
Indeed, this volume equals to the volume of one particle, which is 
$\frac{4\pi}{3}a^3$ times the total number $\mathcal{N}(D)$ of the small 
particles in $D$. One has $\varkappa>0$, so 
$\mathcal{N}(D)\frac{4\pi}{3}a^3=O(a^{-(3-\varkappa)}a^3)=O(a^{\varkappa})\to 
0$ as $a\to 0$.

The limiting effective field $E_e$ satisfies 
equations \eqref{eq76} and \eqref{eq78}, but does not,
in general, satisfy the equation 
\begin{equation}
\label{eq80} \nabla \cdot E_e = 0. 
\end{equation} 
Indeed, the solution to \eqref{eq76} is unique, and
it follows from \eqref{eq78} that
\begin{equation} \label{eq81} \nabla^2 \eta 
+\mathcal{K}^2\eta=-\nabla \mathcal{K}^2(x)\cdot
E_e, \quad \mathcal{K}^2(x):=k^2+C(x), 
\end{equation} 
where $\eta:=\nabla\cdot E_e$.
Since \eqref{eq81} is inhomogeneous,
and $\nabla \mathcal{K}^2 \cdot E_e\neq 0$, in general,  
its solution $\eta\neq 0$, in general.

The unique solution to \eqref{eq76} does not satisfy, in general,
the Maxwell's equation \eqref{eq7}. 
Thus, in the limit $a\to 0$ one gets a vector field $E_e$ which 
satisfies a new equation \eqref{eq76}.

What is the physical meaning of this new equation \eqref{eq76}?
This equation is equivalent to equation \eqref{eq78} 
the solution of which is required to satisfy the radiation condition.
If one rewrites equation \eqref{eq78} in the form:
\begin{equation} \label{eq82}
\nabla\times\nabla\times E_e=\mathcal{K}^2(x)E_e+\nabla \nabla \cdot 
E_e,
\end{equation}
then one can interpret the term $\nabla \nabla \cdot E_e$
as the "current" $i\omega \mu J$ in the Maxwell's equation.
Indeed, Maxwell's equations 
$$\nabla\times E=i\omega\mu H, \quad \nabla\times
H=-i\omega\epsilon'(x)E +J,$$
imply
\begin{equation} \label{eq83}
\nabla\times\nabla\times E=K^2(x)E+i\omega\mu J.
\end{equation}
Therefore the term $(i\omega \mu)^{-1}\nabla \nabla \cdot E_e:=J$ can be 
interpreted 
as a "non-classical" current: the classical current is
of the form $\sigma E$, where $\sigma=\sigma_{ij}$ is the conductivity 
tensor.    
In our interpretation tensor $\sigma$ is a differential operator:
$\sigma=\sigma_{ij}=\frac {\partial^2}{\partial x_i \partial x_j}$,
so
$\sigma E=\nabla \nabla \cdot E$.

A different interpretation of this term may be based on 
the assumption that the vector of the induction $D_e$
in the limiting medium is related to $E_e$ non-locally:
$$D_e(x)=\tilde{\epsilon}E_e(x)+\int_D\chi(x,y)E_e(y)dy,$$ 
where $\chi(x,y)$, the susceptibility tensorial kernel, 
is the kernel of the non-local operator defining the polarization vector 
in the limiting
medium. Such non-local operator of polarization will produce the term 
$\nabla \nabla \cdot E_e$
if $\chi(x,y)$ is a distributional kernel of the form
$\chi(x,y)=(\omega^2 \mu)^{-1}\nabla_x \delta(x-y) \nabla_y $.

Indeed, Maxwell's equations in this case are:
$$\nabla \times E=i\omega \mu H,\quad \nabla \times H=-i\omega 
\tilde{\epsilon}E -i\omega\int_D\chi(x,y)E(y)dy.$$
With the above choice of $\chi(x,y)$, these equations imply:
\begin{equation} \label{eq84}
\nabla\times\nabla\times E=\mathcal{K}^2(x)E+\nabla  \nabla \cdot E.
\end{equation} 
This is equation \eqref{eq82}, which is equivalent to
\eqref{eq78}. The function 
$\tilde{\epsilon}$ in the above Maxwell's equations is related to
$\mathcal{K}^2$ by the usual formula $\omega^2 \mu \tilde{\epsilon}=
\mathcal{K}^2$.

It would be of interest to understand the properties of the 
limiting medium from an experimental point of view.

\section{Spatial dispersion and negative refraction\label{s4}} 
If
$n^2(x)=n^2(x,\omega)$, so that $\omega=\omega(\mathbf{K})$, then one says that
spatial dispersion takes place. Here $\mathbf{K}$ is the wave vector,
$|\mathbf{K}|=K$. Its direction is given by the unit vector
$\mathbf{K}^0$, which is oriented along the direction of the propagation of the 
wave front.
The group velocity $v_g$ is defined by the formula
$$ v_g=\nabla_{\mathbf{K}}\omega(\mathbf{K}), $$ while the phase velocity 
$v_{ph}$ 
is defined by the
formula $$ v_{ph}=\frac{\omega}{|\mathbf{K}|}\mathbf{K}^0,\quad
\mathbf{K}^0:=\frac{\mathbf{K}}{K},\quad K:=|\mathbf{K}|. $$ Negative
refraction means that the group velocity in a medium is directed opposite to
the phase velocity.

A sufficient condition for an isotropic material to have negative refraction 
was derived in \cite{[R533]}.
Namely, in an isotropic medium $\omega=\omega(|\mathbf{K}|)$, and
$$
\omega n(x,\omega)=cK,\quad K=|\mathbf{K}|,
$$
where $c>0$ is the wave speed in the free space. Applying the gradient
with respect to $\mathbf{K}$ to
the above equation yields
\begin{equation}
\label{eq85}
\big{[}n(x,\omega)+\omega\frac{\partial n}{\partial \omega}
\big{]}\nabla_{\mathbf{K}}\omega(|\mathbf{K}|)=c\mathbf{K}^0.
\end{equation}
If $\Image n(x,\omega)=0$, then \eqref{eq85} shows that 
$\nabla_{\mathbf{K}}\omega(\mathbf{K})$ is directed opposite to 
$\mathbf{K}^0$ if and only if
\begin{equation}
\label{eq86}
n(x,\omega)+\omega\frac{\partial n(x,\omega)}{\partial \omega}<0.
\end{equation}

To create a material with negative refraction it is sufficient to create 
an isotropic material with a refraction
coefficient $n(x,\omega)$ satisfying inequality \eqref{eq86}. This can be 
achieved by choosing the constants
$c_{1m}$ in \eqref{eq68} depending on $\omega$ so that the function 
$C(x,\omega)$ would generate by the formula
\begin{equation}
\label{eq87}
n^2(x,\omega)=1+k^{-2}C(x,\omega)
\end{equation}
the refraction coefficient which is a real-valued function satisfying 
inequality \eqref{eq87}.

\section{Conclusions\label{s5}}
A theory of electromagnetic (EM) wave scattering by many
small bodies is developed. A  rigorous 
reduction of the many-body scattering problem
to solving linear algebraic system is given bypassing 
numerical solution of the integral equations.
An equation for the effective field in a medium consisting
of many small particles is derived. It is shown that by choosing 
large number of small inhomogeneities with special properties one can 
create a medium with a desired refraction coefficient.

\end{document}